# Quantifying the selective, stochastic, and complementary drives of the institutional evolution in online communities

Qiankun Zhong[1,*], Seth Frey[1], Martin Hilbert[1]

[1] Department of Communication, University of California, Davis, Davis, USA

[*] Direct correspondence to Qiankun Zhong, (857)218-0274, qkzhong@ucdavis.edu, 469 Kerr Hall, UC Davis, Davis, CA 95616

Institutions and cultures evolve adaptively in response to the current environmental incentives, usually. But sometimes institutional change is due to stochastic drives beyond current fitness, including drift, path dependency, blind imitation, and complementary cooperation in fluctuating environments. Disentangling the selective and stochastic components of social system change enables us to identify the key features to organizational development in the long run. Evolutionary approaches provide organizational science abundant theories to demonstrate organizational evolution by tracking particular beneficial or harmful features. We measure these different drivers empirically in institutional evolution among 20,000 Minecraft communities with the help of two of the most applied evolutionary models, the Price equation and the bet-hedging model. As a result, we find strong selection pressure on administrative rules and information rules, suggesting that their positive correlation with community fitness is the main reason for their frequency change. We also find that stochastic drives decrease the average frequency of administrative rules. The result makes sense when explained in light of evolutionary bet-hedging. We show through the bet-hedging result that institutional diversity contributes to the growth and stability of rules related to information, communication, and economic behaviors.

**1. Introduction**

What are the main reasons that drive institutional changes in organizations? This is a central question across organizational theories. Indeed, the major reasons that lead the institutional changes can often be taxonomized by the type of answer they provide: internal stability, external pressure, information transmission, institutional isomorphism, path dependency, and so on. Some of those reasons are directly related to the payoff of implementation of the institutions; whereas some are driven by stochastic forces. While stylized facts and intuition abound in this area, we have little empirical evidence, which hinges both on the availability of adequate data, and on general frameworks for comparing these sources of change, and showing how they work together.

The migration of organizations to digital platforms allows researchers to obtain more adequate data, thanks to the digital footprint online organizations inevitably leave behind. Traditionally, it was very difficult to obtain statistically significant samples on comparable organizations. Lab experiments with a large N of communities is expensive and nearly impractical to run. Natural experiment cannot ensure similar enough samples to accurately infer the effects of the variable. Digital trace data on online communities make it possible to monitor the intergenerational frequency changes in rules (1), because it provides fine-grained information on exactly when rules are implemented, changed, and removed among thousands of similar enough online communities

As for adequate frameworks for the quantification of the resulting dynamics, the evolutionary framework adopted by researchers in various social science disciplines, including communication (2–4), economics (5–8), and sociology (9,10), provides theoretical and methodological support to answer this question. In the past few decades, social science researchers have used concepts from biological evolution as an analogy to characterize four main stages of institutional development: variation, selection, retention, and struggle (11). This framework categorizes the various institutional changes by the mechanism that drives them and provides explanations from the perspectives of both the organization and the environment. At the same time, the evolutionary framework provides adequate tools to help us represent this analogy with mathematical relationships and explain the macro-dynamics based on a few first principles in given conditions (12). With empirical data, evolutionary models allow for quantifying the strength of different drives behind institutional changes and predicting future development.

One of the most comprehensive and successful models to describe the biological evolutionary process is the Price equation (13). The Price equation partitions total evolutionary changes into two components: deterministic changes driven by natural selection and stochastic changes driven by all other forces, including adaptation, mismatch, drift, and biased transmission. The Price Equation thus provides mathematical tools to separate selective forces and stochastic forces and reconcile different sources of change in a community(14,15).

At the same time, the digital trace data on online communities make it possible to monitor the intergenerational frequency changes in rules (1), because it provides fine-grained information on exactly when rules are implemented, changed, and removed among thousands of similar enough online communities. In this study, we take the advantage of online community data and monitor rule changes among 20,000 Minecraft communities over two years. Online platforms including Wikipedia, Reddit, and Minecraft, provide a great opportunity to study the intergenerational changes of the modular institutional traits among thousands of small-scale communities.

Using the Price equation, we are able to quantify how much of community fitness is derived by natural selection and how much by stochastic forces that are not directly related to the success of the communities. We can also explore further in the institutional structure among online communities to ask whether evolutionary forces are difference among different types of rules. For example, are rules facilitating centralized top-down communication driven by more selective forces compared to rules promoting more decentralized interpersonal communication? Are stochastic forces more prevalent in rules regulating user behavior compared to rules regulating administrative behavior?

Furthermore, is this result robust to the changes in the environment? The bet-hedging method provides us a tool to use the information from the environment to match the frequency of rules, which produces a benchmark for the theoretically optimal strategy of rule distribution. The bet-hedging method thus allows us to ask two questions: First, is the frequency change in one type of rules caused by this rule type only or is it also influenced by other types of rules? Second, is the optimal distribution of rules consistent the Price equation result? If the two models produce consistent result, we can conclude that the selection and stochasticity calculated through the Price equation are robust against the changes and uncertainty in the environment; Otherwise, we expect that the environmental changes play a bigger role in the evolutionary dynamics of institutional changes.

As a result, we found that there is strong selection in the Minecraft environment that drives the frequency changes of some rules, while at the same time, drift also exist among administrative rules that reduces their frequency. At the same time, the bet-hedging result suggests that the communities need to subsidize other types of rules to build resilience against environmental fluctuation.

*1.1 Institutional Change*

The development of institutions has been a key research aspect of organizational studies(16–18). To understand why and how institutions change, social science disciplines including communication (3), sociology (10), and economics (7) have adopted an evolutionary framework to understand the dynamics of institutional development. Institutions as a set of rules to constrain behavior can be transmitted via communication processes and social learning (17,8). The evolutionary approach to institutions allows us to examine both the processes involved in the origin, maintenance, and spread of specific rules as well as the complex ways different rules can interact to produce emergent properties at the populational level.

### 1.1.1 The arguments for adaptive selection

The evolutionary framework in organizational studies focuses on the natural selection over rules. The selection of rules is a process by which rule frequency increases or decreases as a result of the direct payoff contributed by the implementation of rules. All selective processes are characterized by variation, heritability, and competition (19,11). In an institutional context, variations of rule arise across groups and with this variation. With the variation and differential payoff of the rules, there should be also some forms of competition between the institutions on how they are beneficial for achieving organizational goals in economic growth (20), political stability (21,22), successful localized management of common-pool resources (23), and long-term resilience (17,24,25). Selective forces over institutions can occur in three conditions. First, groups with high-payoff institutions outcompete other groups, replacing those groups or imposing their institutions on them (26). For example, the rise of information and communication technologies (ICTs) has enabled a shift from group-based societies to network-based societies because the latter gained a higher benefit with ICTs (27); Second, group members have high leverage to migrate to communities with better institutions at a low cost (28). Banzhaf andWalsh (2008) provided empirical evidence that supports the notion that households "vote with their feet" for better institutions that promote environmental quality; Third, certain institutions are more likely transmitted from one group to another. Zhong and Frey (2020) found that centralized rules are more likely to be transmitted than decentralized rules between online communities with overlapped membership.

### 1.1.2 The arguments for stochasticity

Although social science research that adopts an evolutionary framework mostly focuses on natural selection, many institutional changes are not driven by selective forces. Those non-fitness-related changes are categorized as drift or stochastic forces (7). Two major mechanisms in organizational research characterize this type of change in institutional settings: path dependency and institutional isomorphism. Path dependency refers to the process by which institutional development depends on a unique series of past events. The path cannot be retracted, nor can it be easily deflected on. Path dependency can be explained through diverse mechanisms, including self-reinforcement (31,32), positive externalities (33), and lock-in (34). Although institutional changes driven by path dependency can be beneficial for organizational success, for the time being, the increased frequency is not related to organizational success. Institutional isomorphism (35) refers to the process that organizations borrow routines, rules, and behavior from other organizations regardless of the possible mismatch between the adopted institution and the organizational context. Institutional isomorphism is explained through the organization's internal bounded reality and the uncertainty or pressure of the external environment (36–38).

### 1.1.3 Integrating and disentangling selective forces and stochastic forces

But even if natural selection is overemphasized for explaining organizational change, the evolutionary framework has major benefits for studying institutional and organizational development. Among other things, it provides a formal theoretical framework based on first principles about how inherited traits will change over time given certain conditions. It is precisely these tools that let us articulate the relationship of natural selection to the many other evolutionary processes at work in social system change. In social sciences, evolutionary explanations are often conflated with the selective processes. Stochastic processes, although well studied in organizational studies, are usually not considered from the perspective of institutional organizations. The separation of the two main forces leads to some problems in identifying the true mechanisms of institutional changes. For example, institutional development driven by path dependency may also have direct benefits that are selected for in the competition with other institutions. For example, the rise of platforms, including Apple's iOS and Google's Android, gained both builder and developers benefits at the beginning. But the lock-in benefits gained from the platform discourage the construction of gateways, and thus it forces developers to commit to just one platform or to build and maintain multiple versions of the same product (39). On the other hand, rules that can help achieve institutional goals may also be borrowed by other groups blindly without considering the context. Lowrey found that although there is blind isomorphism among the partnership between newspapers and TV stations, the level of partnering is predicted by concrete benefits and availability of resources (40). These examples show that the selective and stochastic forces are often conflated in institutional development. Focusing on one side of the story cannot provide a full picture of how different mechanisms work together in institutional changes. If we can integrate different institutional change mechanisms, we will be able to answer one of the most important questions through this integrated evolutionary approach: how can we tell apart the selective and stochastic forces in institutional development. In other words, how do we know whether the rule frequencies increase or decrease for their contribution to the organizational goals or for other reasons including path dependency and institutional isomorphism?

It has been difficult to answer this question empirically. First of all, institutions and other social-environmental processes, especially culture, are all endogenous processes. It is not straightforward to establish causal links between institutions and other factors (41–43). Second, the evolutionary processes can be separated into discrete phases analytically, but they are often linked in continuous feedback loops, making it difficult to map evolutionary stages in theory to empirical data (11). Variation provides sources for selection, but the selected traits after transmission and retention will in turn reduce variation among populations. As a result, the evolutionary process cannot help us decide any moment in the process but rather forms a dynamic system driven by different evolutionary forces. Third, selective and stochastic forces can vary across time. Institutions that have been beneficial at the early times can end up reducing the growth speed (e.g., the lock-in effects). It requires both a clear identification strategy and longitudinal data to calculate their time-variant and average strength. Last, it is difficult to quantify institutions and institutional changes due to their complex natures. There is no clear definition that decides whether two institutions are comparable or whether we should take into account the interactions between rules within one institution.

In this paper, we address those difficulties by using the Price equation and longitudinal data on Online communities to disentangle different mechanisms in institutional evolution. We use the longitudinal online community dataset to make quantitative comparisons on institutions between thousands of organizations and apply the Price equation as a statistical strategy to make clear estimation of selective and stochastic forces.

*1.2  The Price equation*

The Price equation (13) is one of the best-known biological evolutionary models with wide applications. It is a theorem that represents any system of differential transmission (44). In its original form in population biology, the Price equation provides a way to understand the effects that gene transmission and natural selection have on the frequency of alleles within each new generation of a population. Due to its abstract mathematical articulation, the Price equation is applied broadly in anthropology and economics (14). With the evolutionary framework we illustrated in organizational studies, it is reasonable to also apply the Price equation to organizational studies. The Price equation partitions total evolutionary change into two components: an abstract expression of natural selection (selective forces) and all other evolutionary processes (stochastic forces). The two pieces of the Price equation together can represent multiple evolutionary forces such as natural selection, shift, and biased cultural transmission. One most general partition of Price Equation is

$$\Delta z \ = \ COV[\frac{w_i}{w}, z_i] \ + \ E[\delta_i]$$

In which $w_i$ refers to the direct fitness-related change in community i associated a cultural trait. In situations where we can draw direct causal between the cultural trait and the change in fitness, $w_i$ can be interpreted as the payoff of the cultural trait. $z_i$ refers to the frequency of the trait in community i ,and $\delta_i$ refers to the random change of the trait frequency in community i. This equation establishes that the fitness-correlated selective forces ($COV[\frac{w_i}{w}, z_i]$) and the fitness-uncorrelated, stochastic forces ($E[\delta_i]$) contribute together to the frequency change of a cultural trait $\Delta z$.

With the theoretical mapping from cultural biological evolution to institutional evolution, we can use the Price equation to estimate the selective and stochastic forces in institutional changes at the level of rules.

### 1.3 Online Communities

Longitudinal data from online communities make it possible for us to quantify institutional changes and extract the measures to apply the Price equation empirically.

Online platforms including Wikipedia, the discussion platform Reddit, and the game Minecraft offer a meta-population of online communities. This type of large-scale groups of communities makes it possible to compare the institutions of thousands of communities within the same macro environment and cultural context (45,46). The communities within the same platform often face the same collective action problems and pursue the same organizational goals, allowing for meaningful comparison of institutions.

In recent years, through methods including API and webscraping, we have been able to acquire longitudinal data on online communities and study the long-term institutional development of many thousands at the same time (45,47–51). In this research, we monitored over 20,000 Minecraft servers, which allow various user activities including building with blocks, gathering resources and interacting with each other. The servers thus function as communities users can engage with. The Minecraft environment hosts millions of communities that compete for the scarce physical and virtual resources and struggle for the same organizational goal — to recruit and retain members. The same collective problems and goals they are facing put them under selection pressure, whereas the various choice administrators and community members have granted them space for stochastic drift. We collected data on the rules each community implements over two years. The modular rule sets, which are called "plugins" in the Minecraft world, provide a standard measurement to quantify institutions and set the unit of analysis at the rule level. The plugin types can then be used as a measurement for institutional traits. By calculating the frequency change and variance of one type of plugins, we are able to apply the Price equation in institutional settings.

Using the Price equation and online community data, we try to answer:

*RQ1: What are the selective and stochastic forces that drive frequency changes in different kinds of rules among online communities?*

*1.4  Time variance and institutional diversity*

Environmental fluctuation exogenous to culture and institutions have a large influence on cultural and institutional evolution. The frequency and intensity of environmental changes affects which type of cultural and institutional trait is selected and stabilized in the long run. For example, Roger's model explains how conformity evolves only in situations where environmental changes are not frequent (52). Giuliano and Nunn use a set of historical data and validate that populations that experience more cross-generational temperature instability attribute less importance to traditional values (53). Richerson and Boyd attribute the emergence of cumulative culture to climate change in the late Pleistocene (12).

In the case of Minecraft, the software environment and version change may cause changes in the payoff of implementing one type of rules and influence its evolutionary trajectory. The environmental influence in Minecraft can thus operate on both the selective forces and stochastic forces. For example, when the overall online community environment becomes more unpredictable or unstable, it is possible that institutions with decentralized rules that promotes peer interactions are more likely to be selected for comparing to centralized rules that reinforces top-down hierarchies (54). At the same time, the uncertain environment may increase blind imitation (35) and leads to stochastic institutional changes. Thus:

*RQ2: Is the evolution trajectory of rules influenced by environmental changes among online communities?*

So far, we consider how single institutional traits (rules) evolve in various environments. However, oftentimes organizational development relies on complementary rules functioning together. Ostrom proposed the Institutional Analysis and Development framework to analyze various social institutions and provided empirical evidence the contribution of institutional diversity on robust self-organized institutions (23). Page provided evidence that supports the benefits of diversity in complex systems, especially in response to external shocks and internal adaptations (25). In Minecraft, we have four types of meaningful rules related to governance. However, when we zoom in and only focus on a single type of rules, the Price equation forces us to include the influence of other types of rules in stochastic forces and environmental factors. Whether other types of rules can interact with one particular type to operate together on institutional evolution through rule diversity requires further analysis. Motivated by Ostrom's and Page's theory, we ask:

*RQ3: Is the evolution of a single type of rules influenced by rule diversity among online communities?*

**2. Materials and Methods**

*2.1 Data*

We collected longitudinal plugin implementation data from 370,000 Minecraft servers through API queries bihourly between Nov 2014 and Nov 2016. After filtering out servers that were disconnected for the duration of data collection (~220,000), those that did not survive for at least a month (~70,000), and those that did not report full governance information (~75,000), we end up with a sample of 14,859 servers (we address the limitation resulted from this data deletion process in the Limitations).

In Minecraft, plugins are modular programs administers rely on to manage the servers. These plugins are modular programs that administrators can install on their servers to automatically implement rules and other institutional constructs (See appendix A for detailed descriptions on plugins). In the digital world, code is law (55) By mixing and matching plugins, Minecraft server administrators establish formal institutions to maintain community survival and achieve community success. The Minecraft community has developed almost 20,000 plugins listed under 16 categories, among which Frey and Sumner concluded 4 rule types directly related to governance: top-down administration, information broadcasting, communication, economy (56). Administration rules enhance administrators' control over community and user behavior; Informational rules facilitate information sharing from administrators to users; communication rules improve communication between players; economic rules protect private property and enable trades. To quantify institutional changes and analyze the evolution process in Minecraft, we used this classification to categorized the plugins. To fit the Price Equation, we summarize community-level data at the unit of one month. As the median "lifespan" of a server is 9 weeks, this aggregation provides an appropriate timescale to capture the dynamics of intra- and inter-generation cultural transmission.

We take a community as an organism and the share of rules (plugins) as the institutional trait or cultural variant it displays. In this large group of communities in Minecraft, we see the different communities exhibit different institutional traits (cultural variants), which constitute the overall distribution of institutional traits in Minecraft.

In Minecraft, we do not know if after a community dies, the governing knowledge is inherited by its members to pass on to the next generation. In this sense, we do not strictly follow a genetic inheritance model. However, when a new community starts, to maintain community survival and deal with collective action problems, the community administrators need to learn socially from other communities or learn independently from the environment to establish their institutional traits or governing style. The process of learning can be seen as cultural transmission that changes the overall distribution of rule shares.

Different forces contribute to the distribution changes. Selective forces act on Minecraft communities in two ways. First, communities that employ the governance style beneficial for community survival will last longer. For example, if administrative rules are the most beneficial for community survival, communities that employ a large share of administrative rules will last longer. In contrast, communities that employ a governance style with a small share of administrative rules will die out faster. This differential survival rate of different governing styles will lead to a shift in the overall distribution of administrative rules. Second, communities that employ the governance style beneficial for community success are more likely to be copied by other communities. The spread of successful governance styles can also change the overall distribution of rule shares.

Stochastic forces also act on Minecraft communities mostly in two ways. First, communities learn from other communities blindly. When the learning is not led by success bias but rather by proximity or uncertainty, this type of copying will lead to drift in the overall distribution. Second, when players cultivate cultural preferences of specific governance styles, they are likely to spread these specific rule shares to other communities they migrate to (30).

Additionally, mutation provides additional variation for selection. In Minecraft, the introduction of new plugins or individual learning to establish new governance styles can be seen as mutation.

Administrators do not know whether their implementation of one type of rules instead of other types of rules is due to selective or stochastic forces. We as well cannot accurately identify from the interactions between communities the selective forces or stochastic forces. What is available in this dataset is the collective pattern of rule distribution. The advantage of using the Price equation is that, in this case, it isolates the selective forces from the stochastic forces statistically without specifying a large number of possible mechanisms of each basic process.

*2.2 Price Equation*

The Price Equation provides a powerful generalization of the forces contributing to evolutionary changes in institutional structures — analogous in our framework to biological traits — among online communities — which correspond to individual organisms, the population of communities corresponding to the population of organisms. Here we demonstrate one possible decomposition of the Price Equation that focuses mainly on distinguishing between the strength of the correlation between the relative growth of the population ('fitness') and the presence of a certain community trait ('selection' based on that trait), and the strength of other stochastic fluctuations.

Consider the Minecraft environment consisting of multiple servers each indexed by $i$ ($N = 13,859$ servers). Within each community we identify rules of four categories, administrative, informational, communication, and economics rules (see Figure 1a). Taken administrative rule as an example, the relative frequency of administrative rules in community i is $z_i = r_i/R_i$ where $r_i$ is the number of administrative rules and $R_i$ is the total number of rules in community i. Some communities might have many rules, but a small population (see Figure 1b), so we consider it useful to create a measure of rule fraction weighted by membership population size. In most cultural evolution work, to estimate the fraction of a cultural trait within a population, researchers would either calculate the number of individuals that carry a specific cultural trait weighted by the overall population size (57) or the number the artifacts with a kind of cultural feature weighted by the production size(58). It is a bit tricky to do this kind of calculation in organizational and institutional evolution because conceptually, the individuals in the organization do not directly carry the cultural trait and the total number of rules do not reflect the "production size" of the community. Therefore, to establish the connection between the rule frequency to organizational size and the make sense of differential cultural transmission, we use the frequency of administrative rules weighted by population to estimate the fraction of the institutional trait among all communities. For simplicity, we refer to this measure as the reach of rules (their contact with population).

We calculate the mean reach of administrative rules across all communities $m = \sum_i p_i z_i$, where $p_i = n_i/n$ is the relative membership size $n_i$ of community i over the total amount of active population $n$ among all communities, and $z_i$ is again the relative frequency of administrative rules over all rules in community $i$ (see Figure 1c).

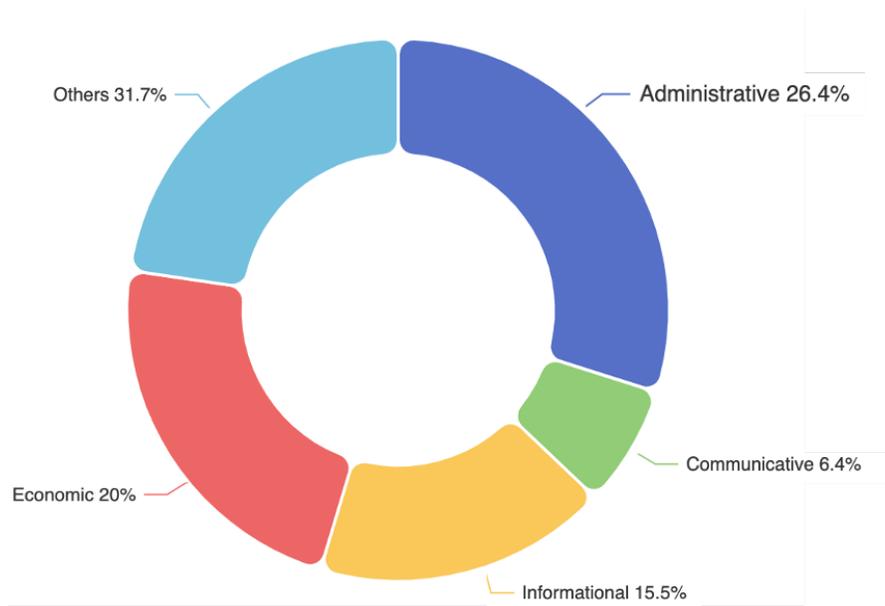

**(a) The rule fraction $z_i$ of a community $i$ at time $t$**

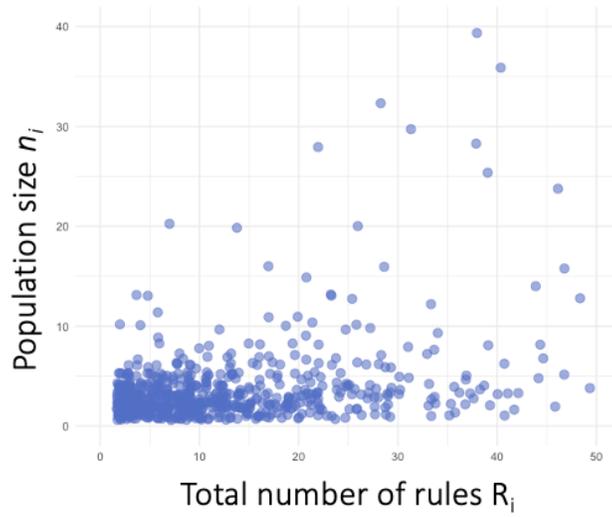

(b) **Membership size ($p_i$) by total rule number ($n_i$) scatter plot at time $t$**

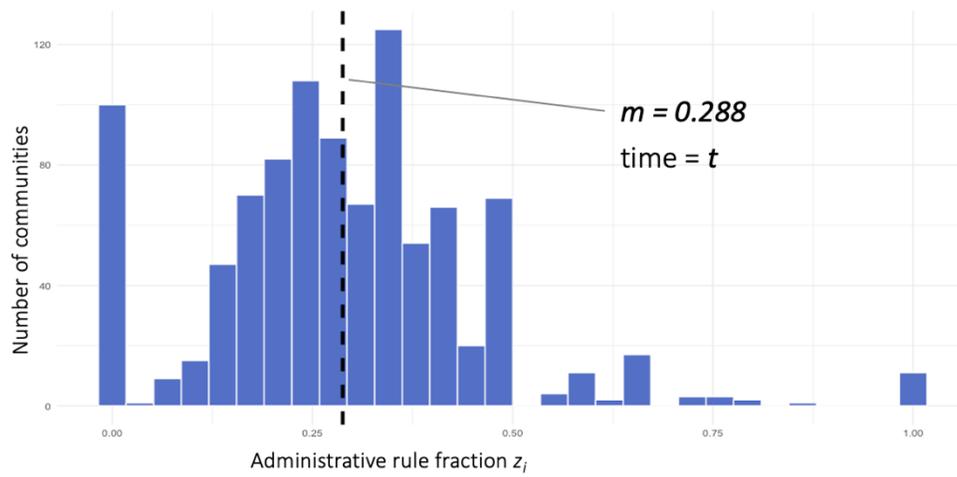

$m = 0.288$

time $= t$

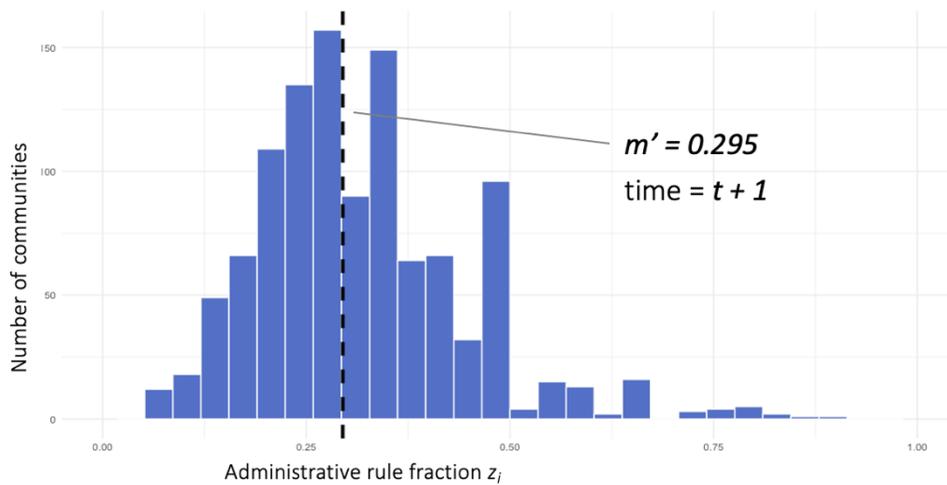

$m' = 0.295$

time $= t + 1$

(c) **Histogram of the administrative rule fraction ($z_i$) changes from time *t* to to time *t+1***

*Figure 1* **Model setup (a) Rule share pie chart of a sample community *i* at time *t*:** each community has a fraction of administrative rules ($z_i$); **(b) Membership size ($p_i$) by total rule number ($n_i$) scatter plot at time *t*;**(c) Histogram of the administrative rule fraction ($z_i$) changes from time *t* to to time *t+1*, which also changes average population reach *m*.

The fitness of the reach of administrative rules (tracked with the letter $m$) is therefore dependent on the differential growth of certain kinds of rules (tracked with $z_i$) and the population size of the respective community that uses these rules (tracked with $p_i$). The Price equation decomposes the change in the reach of rules (change in $m$) into how much the presence of these rules covary with the growth of the population (a strong positive covariance would detect that more of this kind of rules goes together with an increase of the population size) and into all remaining stochastic fluctuations observed in the number of this kind of rules. This allows us to quantify how strongly the presence of a certain kind of rules covaries with change in population size, plus a remaining stochastic term.

In this sense, we consider two sources of change in the frequency of administrative rules. The first source is dependent on differential membership population growth (fitness) associated with the different share of administrative rules within each community. This derives the covariance term of the Price equation. In Minecraft context, we aim to covary the differential membership growth rate in a community with a specific share of administrative rules. For example, if administrative rules are related to a higher population growth rate within a community, we may see that communities that implement 100% administrative rules have a higher-than-average population reach growth. Several mechanisms may lead to this relationship.[1]

We now introduce a new variable, $w$, to track the rate of change of population shares, where $p_i'$ is the proportion of administrative rules in the next time interval. .

$$p_i' = p_i \frac{w_i}{w} \quad (1)$$

The mean rate of change will then be $w = \sum_{i=1} p_i w_i$. When the size of a population increases, the relative reach of the used rules increases. In Minecraft communities, community goals are to survive and to recruit and retain more members, $\frac{w_i}{w}$ tracks the relative fitness change of communities. This source of change in the reach of the administrative thus can be seen as fitness-related change and is tracked by expanding equation (1) with the static number of rules:

---

$$p_i' z_i = p_i \frac{w_i}{w} z_i \quad (2)$$

When the increase rate of the membership population is greater than the mean growth rate (i.e., $\frac{w_i}{w} > 1$), the population reach of administrative rules in community i mi increases without the frequency change of administrative rules within the community $z_i$. This fitness-correlated process is conceptually equivalent to selection acting at the scale of organizations.

It is worth noting that this is not a perfect replicator at the individual level in the Minecraft context because communities may start (birth) or go offline (die) during the time period we collect data. However, at the group level, when new communities come online and copy those are successful, the fraction of active populations that are constraint under the same rule strategy increases. When communities get offline, it does not only lose its own share in the population that are governed by its rule strategy but also do not provide source for other communities to copy anymore and thus will reduce the population share of this type of rules. Thus, even without perfect individual-level replicator, the cumulative institutional changes at group level remains the same.

A second source of change in the weighted frequency of administrative rules is stochastic fluctuation, which may arise from drift(7) or transmission errors(59). In this case, we can write

$$p_i z_i' = p_i (z_i + \delta_i) \quad (3)$$

Where $\delta_i$ is some small random change in the frequency of administrative rules and $z_i'$ indicate their frequency within community i in the next time interval.

Equation (2) and (3) can then be combined into a single specification that simultaneously accounts for both selective and stochastic forces operating over time:

$$p_i' z_i' = p_i \frac{w_i}{w} (z_i + \delta_i) \quad (4)$$

Change occurs both in the population size of each community and in the frequency of administrative rules. Using the above rules, it is straightforward to derive the Price Equation (Price, 1970).

$$\Delta m = m' - m$$

$$= \sum_{i=1} p_i \frac{w_i}{w} (z_i + \delta_i) - \sum_{i=1} p_i z_i$$

$$= \sum_i p_i \frac{w_i}{w} z_i - \sum_i p_i z_i + \sum_i p_i \frac{w_i}{w} \delta_i$$

$$= \sum_i p_i \frac{w_i}{w} z_i \ - \sum_i p_i \frac{w_i}{w} \sum_i p_i z_i + \sum_i p_i \frac{w_i}{w} \delta_i$$

Price recognized the first and second terms are equal to $E[\frac{w_i}{w}, z_i] \ - E[\frac{w_i}{w}]E[z_i]$, which is the covariance between $\frac{w_i}{w}$ and $z_i$. The third term can be rewritten as $E[\frac{w_i}{w}, \delta_i]$. But since the fluctuation $\delta_i$ has no correlation with $\frac{w_i}{w}$, the third term can also be written as $E[\delta_i]$. The equation thus can be simplified as

$$\Delta m \ = \ COV\left[\frac{w_i}{w}, z_i\right] + \ E[\delta_i] \ \ (5)$$

The Price equation derived (did not postulate) a covariance between the relative population growth $\frac{w_i}{w}$, and the relative frequency of a certain kind of rules, $z_i$. If this covariance is positive, it tells us that a presence of a certain kind of rules goes together with population growth. Since there is no intrinsic directionality in a covariance, this can be interpreted in two ways: the presence of a certain kind of rules helps the population to grow; or, the growth of the population helps the reach of the rules to increase its reach. Both help to increase the reach of the rules within the total population among all communities (measured by $\Delta m$).

The totality of the equation implies that the selective forces (the first term) and the stochastic forces (the second term) contribute together to the frequency change of institutional traits $\Delta z$ (13). For empirical application, this can be reformulated as

$$\Delta m \ = \ \beta VAR[z_i] + \ E[\delta_i] \ \ \ (6)$$

with the product of the coefficient $\beta$ of the relative population growth $\frac{w_i}{w}$ on the frequency of administrative rules $z_i$, and the variance $var[z_i]$ in the frequency of administrative rules. In this equation, the slope reflects the strength of selective forces, and the intercept represents the strength of stochastic forces (58).

The intuition behind the equation can be understood through two levels. At the individual community level, when a community first goes online, it needs to install rules from the pool of plugins. The community administrator can learn the governance style from other successful servers (success-biased learning) or the more popular governance style (frequency-biased learning). The administrator can also try out new governance styles or new plugins developed in the Minecraft community (mutation), and they can also learn from other resources (individual learning). The result is that some of those implementations are beneficial for the community to survive longer for this governance style to retain in the population for longer or for the community to be more successful so that other administrators are more likely to learn from them. It could also be the case that the type of rules the community installed are detrimental to the community's success and thus will lead to a shorter lifespan or less of a success to be copied by other communities. At the population level, the average share of rules that are beneficial for the community's survival and success will increase due to the communities that are sustained in the population or social learning mechanisms.

In an organizational context, the process we described here may seem overly simplified and abstract, especially considering all other third variables that may cause membership population increase or community death. Nonetheless, the Price equation describes the system of institutional evolution in a minimal manner and offers a way to identify and quantify selection or fitness correlated rule change (60). The fitness-correlated process is conceptually equivalent to selection acting at the scale of organizations. It is also worth noting that although we construct those process in the model, it does not establish the causal link between rule frequency change and actively population share change. Instead, the empirical estimation will help us to establish the connection between the two.

As a result, the final linear equation form demonstrates a relationship between rule frequency, variation, and selection: different types of rules have variations between their correlation to population growth. When the relationship of variation to total frequency changes (beta) is negative, it indicates that the administrative rule has a negative correlation to the community membership growth. When communities have many variations in the proportion of administrative rules between each other, the high variation provides sources for selection and thus leads to higher frequency change rate, whereas when there is less variation, the competition will be tight and the rate of change will be smaller. A positive slope indicates positive correlation between rule frequency population growth. It suggests that over time, the communities with higher fraction of administrative rules will increase the population reach of administrative rules. Similarly, a negative slope indicates selection against administrative rules, suggesting a higher proportion of administrative rules will decrease its population reach over time.

*2.3 Bet-hedging and information theory*

We take an inductive approach of Price Equation to estimate the strength of selection from frequency change, which enables us to estimate the fitness-related frequency change of a particular rule through the coefficient $\beta$. However, this approach also forces the growth rate to be fixed throughout time: an average relative growth rate is assumed for each group of rules. This modeling strategy is reasonable in a stable environment, but when the environment is changing, it might not reveal the true dynamics in frequency change due to two reasons: First, the selection forces on one type of rules depend not only on how well this rule does on average (algorithmic mean of the fitness-related growth), but also on whether implementing this rule instead of other rules would lead to community failure in a vulnerable time (geometric mean of the growth)(61). The Price Equation uses the arithmetic mean and thus cannot provide a explanation for changing environments. Second, the selective forces on communities depend not only on the quantity of rules that covary positively with population growth but also on the combination strategy of rules. It could be that some rules are beneficial in some periods, while others are useful in other periods. The average over all periods might suggest favoring one rule over the other, but eliminating the other rule entirely might crash the population in certain periods. In the Price Equation, to estimate the selection over one type of rules, we subsume the influence of other rules within our stochastic forces residual $E[\delta_i]$. It helps us focus on the dynamics on a single type of rules, but it does not explain much how different types of rules can work together to prepare communities through changing environments in a synergistic dynamic over time.

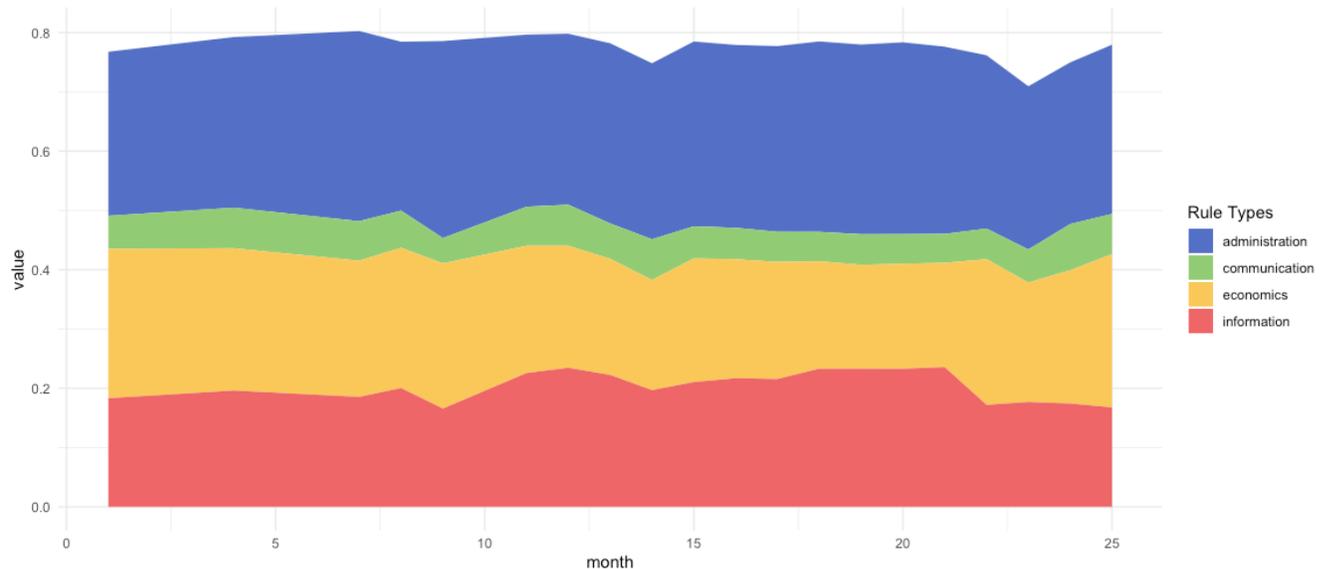

*Figure 2* **Environmental changes throughout time cause changes in the number of communities, but do not seem to change the overall relative proportions of rule types in the population, except for administrative rules**. The Price equation assumes a constant correlation between population growth and the implementation of one type of rules. However, the correlation may vary across time in a fast-changing environment. The changing bandwidth of administrative rules in this figure demonstrates that different rules are influenced differently by environment

One of the methods that can incorporate the two factors not addressed in the Price Equation is bet-hedging. Evolving biological and socio-economic populations can sometimes increase their growth rate by cooperatively redistributing resources among their members. In unchanging environments this simply comes down to reallocating resources to fitter types (62). This would suggest that rules that are not useful for quite some time will get eliminated by natural selection. However, they might become useful later again and their premature elimination would then reduce fitness. Whenever there is a repeating cycle or seasonality in fluctuating environments, it could be useful to restrict the forces of blind natural selection during certain periods to be prepared for subsequent periods (59). For example, it would not be useful to allow natural selection to eliminate all food storage during summer, even if 'blind' natural selection cannot see its utility during periods when the environment provides abundant food sources, as it will be an essential pillar of fitness during the upcoming winter. Neglecting such predictable seasonality and merely working with cross-seasonal averages might suggest that the contribution of the "food storage rule" is on average neglectable over the "go out and forage" rule, and might suggest to eliminate it. However, the population would starve during winter and not make it to the next thriving summer. Any kind of anti-cyclical governmental policy, or any kind of temporal economic subsidy exploits the same logic of maximizing overall fitness in predictably fluctuating environments by combining different kind of institutional mechanisms over predictably fluctuating periods of the evolutionary trajectory (59).

The utility of such portfolio theory-based bet-hedging over time depends on the predictability of the environmental pattern. If the future cannot be predicted, there is nothing to prepare for. However, if a predictable seasonality is known, one can adjust for it. The predictability depends on the amount of information in the pattern, which leads us to information theory.

In technical terms, if the information about the future pattern is completely unrelated to the state of the environment, the mutual information between the cue and the environment is zero. The cue does not help to increase the 'fit' between the evolving population and the environment. At best, a perfectly informative cue would exactly reveal the state of the environment and the remaining uncertainty about the environment would be zero and the population could be adjusted to grow optimally. This can be formalized by the mutual information between the environment and the evolving population: the more you know (about the environment), the more you can grow (the population) (63).

This problem is formalized by portfolio theory and is the basic idea behind bet-hedging. It uses information about in the environment to maximize the long-term increase rate (64). Built upon Kelly's idea, and subsequent expansions (65–68), Hilbert derives a measure to establish cooperative resource redistribution strategy to maximize socioeconomic growth(63). By establishing fitness matrices of rules in different environmental states, we can find out the most efficient rule distribution strategy that allows the sustainability and incrementation of this type of rules. Here we illustrate the bet-hedging method in Minecraft.

First, we need to understand how the environment changes throughout time. The most efficient way to benefit from the changing dynamics in the environment is represented through information theory. Information by definition is related directly to the reduction of uncertainty. The 'mutual information' between a cue about an environmental pattern and a random environmental state measures how much the cue reduces the uncertainty about the state, and can be directly translated in growth potential (67). It turns out that the search for optimal growth consists in the search for the mutual information (or unequivocal signals) between the environment and the evolving pattern (69).

In the case of Minecraft, if we already know the unfolding dynamics of environmental changes,

we then have no uncertainty about the environment and perfect information to predict the state of the environment at any given time. In an ideal world like this, we can quantify how well one type of rules work in different environmental states and make confident decisions about implementing or removing this type of rules accordingly. However, in Minecraft, we have a high-level of uncertainty about the environment changes. What's the most efficient way for us to use the information from previous event to predict the mechanisms of the current environment? Shannon answered this question by calculating the opposite side of information, uncertainty (70). According to Shannon, the likelihood of getting one specific environmental state from all possible environmental states is equivalent to the reduced uncertainty caused by the knowledge on this specific environmental state. We use a probability distribution of whether the environment is friendly for one type of rules to represent the dynamical patterns of the environment. To minimize the number of assumptions to measure the environment and establish one environmental measurement to consistently compare the changes of all four measures, we decide the good or bad state by whether the growth of the centralized rules (top-down administration, information broadcasting) outcompetes the growth of decentralized rules (communication, economics). This is aligned with hypotheses proposed by Perrow that the paradox between centralization and decentralization grows with organizational complexity(71,72). This cut-off allows to fit our data in a binary framework of the computation.

**Table 1** Rule growth rate partitioned by environmental state and rule categories

| State | growth of rule $i$ | growth of other rules |
|---|---|---|
| Good state for centralized rules | $G_1$ | $g_1$ |
| Bad state for centralized rules | $g_2$ | $G_2$ |

To optimize the rule increment in a changing environment, we need to redistribute the rule shares to match the environmental-state probabilities. In extreme cases, rules that don't work best for the environment have a growth rate of 0. The optimal strategy then is to maintain the shares of rule types just as the corresponding environmental-state probabilities. In between extreme cases, we may find that some rules in an unfavorable environment may still have a positive growth rate. This requires us to adjust the rule shares based on both the growth rate of different types of rules and possibilities of different environmental states. Table 1 listed the growth rate different type of rules in a good or bad state, where W refers to the growth rate in a good state for this type of rules and w refers to the growth rate in a bad state for this type of rules.

Solving for the optimal distribution of administrative rules sometimes results in undefined combinations of rule shares, including cases where the optimal share of rule i is negative (d <0) or above 1 (*d >1*). For *d < 0*, it suggests that the optimal strategy is to implement other rules only. Accordingly, for *d > 1*, the optimal strategy is to implement rule i only. The two extreme cases are the so-called "pure strategy". If the optimal strategy for rule i is a pure strategy of full investment on rule i, it indicates that (1) the frequency change in rule i can be attributed to the earlier investment on rule i other than any other rules. In other words, the frequency change of rule *i* is driven by selection solely over rule i; (2) environmental changes do not alter the growth rate of rule *i* (63).

Optimal solution with rule share between 0 and 1 is called "region of bet-hedging"(67), which suggests a mixed proportion of different rules. The conditions on when it is beneficial to take advantage of cooperation among types to outperform blind competitive selection depends on the particular shape of the fitness landscape (59). The more complementary the fitness of types in different environmental states, the proportionally larger the potential benefit of strategic cooperation over competitive selection. If the optimal strategy for rule *i* is in the region of bet-hedging which suggests implementing *p* rule *i* and *1 - p* other rules, it indicates that (1) the frequency change in rule *i* can be attributed to earlier implementation of both rule *i* and other rules. (2) The environmental changes do alter the growth rate of rule *i* and that's why we have to use a mixed rule combination (resources in the space) to deal with the environmental changes (risks in time;73).

### 3. Results

*3.1 The Price equation result*

To match the rule data with membership data, we marked 23 timestamps to estimate the rule changes over time. At each time point for each community, we measured the fraction of each type rules (zi), membership size (pi); We then calculate the average rule proportion weighted by membership size (m) and variation of rule proportion among communities. We partition the Price equation into slope-intercept forms (equation (6). In this equation, the slope reflects the strength of selective forces, and the intercept represents the strength of stochastic forces (58). As shown in Figure 3, each data point refers to a timestamp placed in the coordination of the variation of rule fractions *VAR[$z_i$]* and the average population reach m derived from . We estimate that in communities with administrative rules installed face positive selective forces ($\beta_{admin} = 0.117, p < .001$) and negative stochastic forces ($E[\delta]_{admin} = -3.602, p < .01$). This indicates that, on the one hand, administrative rules have a high positive correlation to community success of recruiting and maintaining members, resulting in a higher probability for this type of rule structure to be learned by other communities. This direct fitness-related benefit contributes to the growth of administrative rules. On the other hand, driven by stochastic factors, including lack of information, cultural preference/resistance, path dependency, or individual learning, administrators tend to reduce the proportion of administrative rules regardless of their direct positive coorelation with community fitness.

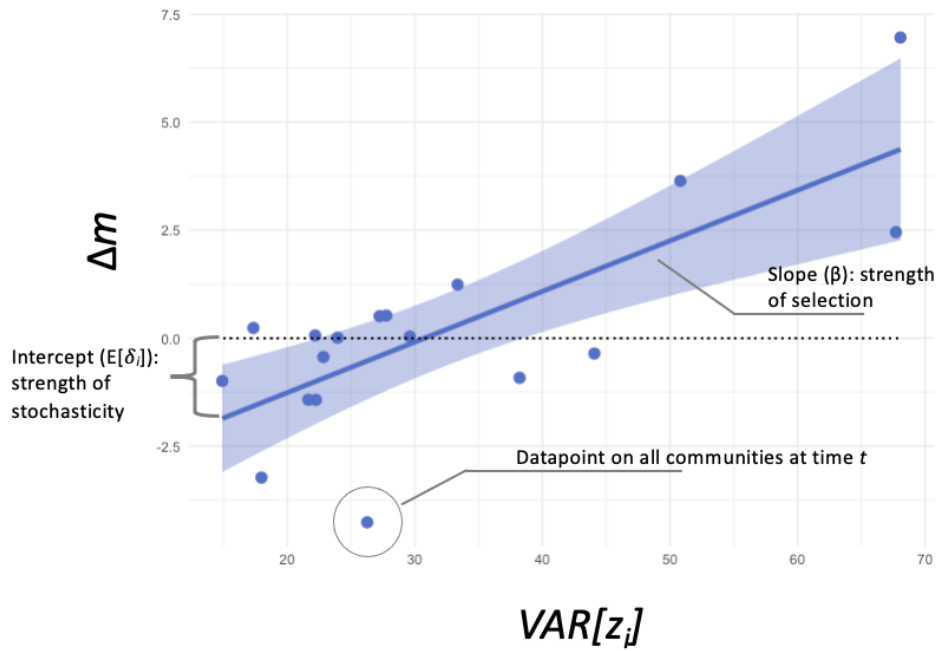

**Figure 3 Communities with administrative rules installed face positive selective forces and negative stochastic forces.** Administrative rules have a positive correlation to community fitness, which also lead to a higher probability for this type of rule structures to be learned by other communities. This direct fitness-related benefit is associated with the growth of administrative rule. On the other hand, other "stochastic" forces including lack of information, cultural preferences, cultural resistance, and random experiment reduce the implementation of administrative rules.

We also find positive selection over information rules ($\beta_{admin} = 0.147$, $p < .05$), indicating that information rules are beneficial for community survival.

We do not find statistically significant selection (slope) or stochasticity (intercept) different from 0 in communication rules and economic rules, indicating the frequency change of communication and economic rules are not significantly different from 0.

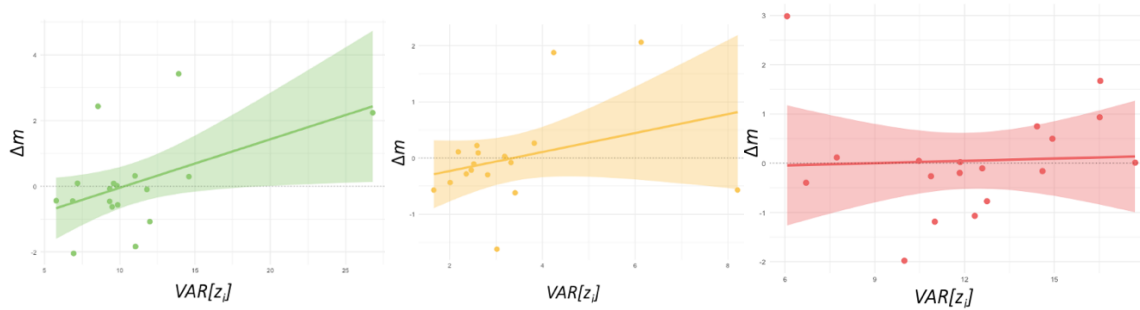

(a) Information                (b) Communication              (c) Economics

**Figure 4 Communities with informational rules face positive selective forces, while there are no effects of communication and economic rules on community prevalence**. We found positive selection over informational rules but not negative stochastic forces (a). At the same time, both selection (the slope) and stochasticity (intercept) in communication (b) and economic rules (c) are not significantly different from 0 throughout time.

### 3.2 Bet-hedging result

We use bet-hedging to validate the Price Equation result and see how the combination of different rules contributes to the rule frequency change.

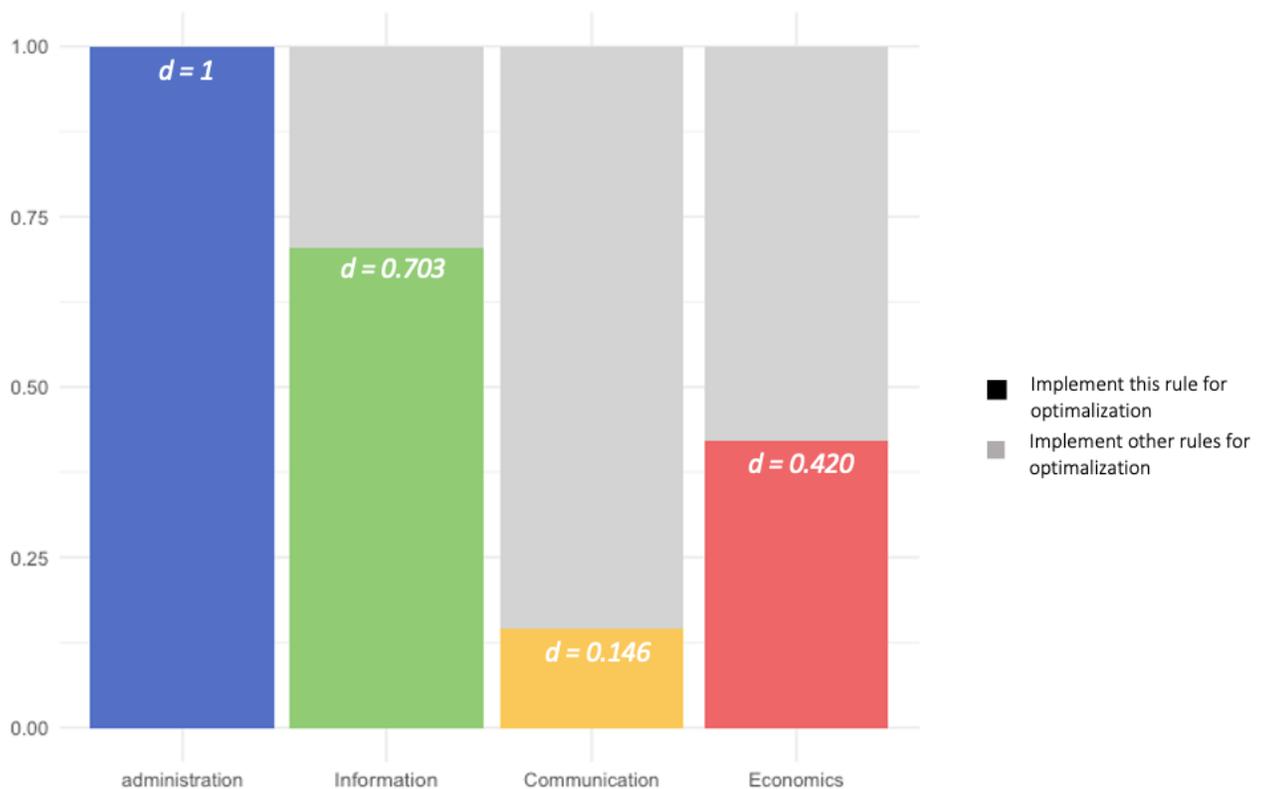

**Figure 5 Most rules show a maximum in their selective effect in combination with the other rule types.** The bars in the figure illustrate the optimal distribution of rule implementation to maximize the growth rate of one type of rule, demonstrating the influence of implementing other types of rules on this type. For information rules to be expressed at a maximum rate in the population, the calculation suggests that they should be implemented with a 30% mix of other rule types. (This is distinct from the question of whether that maximum is positive: whether information rules are positively selected for, as show in Fig. 3). Implementing a mix of rules can help communities survive the period when the direct benefit of information rules are low. As a result, institutional diversity contributes to the long-term growth of communication, information, and economic rules. The optimal distribution of administrative rules, 100%, suggests an absolute strategy for the growth of this most dominant rule type. This may be an artifact of the strong positive selection that communities with administrative rules face, particularly relative to the other rule types. It is also consistent with a conclusion that the correlation between administrative rules and community fitness does not vary as much across time as the other rule types.

The direct result of bet-hedging shows that the optimal distribution of rules for administrative rules to increase is to implement administrative rules only ($d_{admin} = 1$; See Figure 5). In other words, the incrementation of administrative rules can be attributed solely to the earlier implementation of administrative rules. The Price equation suggests that the theoretically optimal strategy is equivalent to the end-result of pure natural selection. As such, it is consistent with the Price equation result that the positive selective force is the only reason for the increase of administrative rules.

At the same time, for information, communication, and economic rules, the optimal share for them is within the region of bet-hedging ($0 < d < 1$; see Figure 5), indicating the environmental changes do alter the growth rateof the three types of rules, resulting in optimal strategies of mixed rule combinations (resources in the space) in response to the environmental changes (risks in time). For these rules, it is useful to keep natural selection at check, as it would overexpose the community to rules which are less favorable in certain kinds of recurring environmental states. There is a "complementary variety" among the diversity of the fitness of these rules in different environmental periods ((62).

Combining Price Equation results, the optimal rule combination for informational rule growth ($d_{information} = 0.703$) shows that although informational rules have a general positive correaltion to community survival and success, this correlation varies throughout time. In a period when the growth rate of informational rules is low, the share of other rules helps the community throughout those difficult times. As for communication rules and economics rules, although they do not have individual selective advantages, they can be subsidized to help communities through environmental changes ($d_{communication} = 0.146;\ d_{economics} = 0.420$).

Overall, we found that the environment for administrative rules is a winner-take-it-all selection-driven situation. But for the other three types of rules, institutional diversity drives rule increment instead of fierce competition and selection. In the long term, it is beneficial overall to maintain a certain mix of rules, even against the elimination pressure exerted by natural selection.

### 4. Discussion

In this study, we use the Price Equation and the bet-hedging method to quantify and isolate the drives of rule frequency changes among online communities. Under the relationships that the Price equation articulates, we found positive selection forces over administrative rules and informational rules. At the same time, stochastic forces including random trials and cultural preferences lead to rule reach decrease in administrative rules. We do not find significant rule reach changes in informational and economic rules. The bet-hedging result of optimal rule share supports this result and provides additional explanation for the stochasticity quantified through the Price equation. We found that the increase in administrative rules is only driven by its positive selection, whereas the increase in information, communication, and economic rules is driven by institutional diversity as well.

The result also provides access to the environmental states of rules. Administrative rules are in an environmental state where competition and selection dominate institutional evolution, whereas, for other rules, diversity and cooperation are the keys to success.

*4.1 Contributions and Implications*

This study used evolutionary frameworks and models to explain institutional development. By using an evolutionary framework, we do not disregard "agency" in institutional changes, but emphasize that in the long run, agency itself becomes endogenous through iterated learning, selection and reproduction of practices and beliefs. On this basis, we integrate the theories in organizational studies and formal models from evolutionary biology to explain the macro dynamics based on first principles in given conditions. The empirical application of the Price equation in this paper helps us quantify selection and stochasticity and thus answers one of the fundamental questions in organizational studies: Are rules and institutions implemented for their direct benefit or for other reasons?

Our approach combines the advantages of comparative studies and mathematical models to show the dynamics and reveal collective patterns of institutional evolutions (24). Through comparative analysis over thousands of communities in the same Minecraft environment, we can control for the spillover effects of other social processes and focus on the frequency change in rules. Through the non-linear mathematical models, we assess the institutional development not as a moment of equilibrium but as an evolving system where changes emerge based on some first principles and stochastic processes. The use of bet-hedging models complements Price equation results, demonstrating a practical application of applying information theory to answer evolutionary questions.

Additionally, our bet-hedging results show the influence of environmental fluctuation in evolutionary processes and point out a path to identify the current environmental state for particular institutional traits. The estimation of the environmental states and their influence can provide valuable information in general risk-avoiding and decision making, especially when other variables are fixed or controlled. This approach loosens up the fixed environment assumptions in evolutionary models and thus helps make more accurate predictions in an uncertain and risky environment.

The bet-hedging results also contribute to the literature of institutional diversity in three perspectives. First, our results support empirically that institutional diversity is beneficial for institutional and organizational development for certain rules. Second, we are able to calculate the boundary conditions of environmental states and specific rules in which institutional diversity have the maximum benefit. Third, we extend the theory of diversity by demonstrating that diversity does not only benefit the overall collective fitness (25) but also contributes to the growth of a single rule (trait).

Although we focus on the online community context, our results, to some extent, can be generalized to real-world communities and provide some implications for policymakers and practitioners. The empirical evidence in this paper suggests that in a fast-changing environment, institutional diversity can be helpful for organizations to build resilience.

Overall, in this research, we join the conversation with the population ecology research of online communities (4,2,74,49,75) to further understand organizational development. Ecological thinking and evolutionary thinking provide two approaches to understanding the frequency change in organizations. In recent years, researchers in different disciplines are trying to bridge the two grand theoretical frameworks and produce more integrated models (76,77). Our work contributes evolutionary thinking to the recent empirical development and advances the development of integrated models and model selection in organizational studies.

*4.2 Limitations*

The Price equation is powerful in explaining the macro patterns of the system but it does not provide direct causal inferences. This is because the Price Equation is ultimately a tautology (14)that describes frequency change. Thus, although we are able to estimate the strength of selection, we do not know what drives frequency changes aside from selection. Anything not directly related to community fitness is concluded in stochasticity, which we cannot explain through the model. At the same time, the Price equation, when applied to cultural and organizational evolution, is difficult to map accurately to organizational activities. In this research, we do not have perfect replicator of rule change mechanisms. Although replicators are not necessary for cumulative, adaptive cultural evolution(78), but it makes estimation and interpretation of the model less accurate than biological evolution estimation. Additionally, we use GLM to estimate selection and stochasticity to guarantee the robustness of the estimator. However, we cannot be sure that the current estimation method may not be the most efficient. It is still debatable which estimation method is the most effective to estimate the slope of selection.

Our application of the bet-hedging method assumes a fixed fitness matrix due to the limitation of the technique (64). Limited by the computation, we can only assume a two-state environment and calculate the shares for the binary rule categories. This limitation simplifies reality and also forces an arbitrary choice of deciding the environmental state. In this paper, we use the relative growth of centralized rules (administrative and informational rules) as the indicator to decide environmental states. This allows us to answer the research question, but at the same time this categorization is still relatively arbitrary and less theoretical. In future work, we may introduce more context-based measures of environmental state based on organizational theories.

Finally, we studied fitness in terms of the reach of rules. We tracked the reach of rules among Minecraft communities for the sake of the influence of their reach, which is a justifiable definition of 'rule fitness' (the selfish rules is propagating). However, it does not tell us anything about some other utility of the rules (e.g. the growth of the satisfaction of users, their enjoyment, the economic or entertainment benefits of rules, etc). This additional step could be achieved by relating the reach of rules to other performance measures of communities, as can be done with structural equation modeling(79).

*4.3 Future work*

Our methods make the first attempt in demonstrating evolutionary thinking in institutional development and also point out where to look into the data when analyzing institutional development. The general contribution is that we show that it is practically possible to apply long-standing formal theories of evolutionary change to calculate concrete and insightful aspects of the evolution of institutions. The digital footprint produced by online communities and organizations allows researchers to advance to this more formal stage of empirical testing and quantification. Existing evolutionary frameworks from evolutionary biology, such as the Price equation and bet-hedging, allow researchers to calculate long-standing measures and interpret them within solid conceptual frameworks.

The Price Equation pointed out where to look at when analyzing other influences aside from selection. For future research, we may want to use this information to look into the influencing factors in stochastic forces. At the same time, the bet-hedging method points out where to look to identify the efficiency of institutional diversity in a changing environment. Future research may narrow the scale of institutional analysis to particular time periods and rule shares to identify institutional effects. Future research can also look into the reasons for subsidizing particular rules.

To summarize, this research highlights the evolutionary thinking of institutional analysis and embraces the opportunity of macro-scale longitudinal analysis on online communities provided by digital trace data. By applying evolutionary models empirically, we are now able to answer fundamental questions of institutional evolution quantitatively and to open the door for future research to study institutions from an evolutionary-system perspective.

**Acknowledgments:** We benefited from invaluable discussions with and comments from Jeff Schank, Sydney Wood, Aviva Blonder, Cristina Moya, Curtis Atkisson and members of the California Workshop on Evolutionary Social Science.